\documentclass[10pt,ioptwocol]{iopart}
\usepackage[colorlinks, linkcolor=blue, anchorcolor=red, citecolor=blue,urlcolor=blue]{hyperref}
\usepackage{graphicx}
\usepackage{epsfig}
\usepackage{float}
\usepackage{epstopdf}
\usepackage{cite}
\usepackage{indentfirst}
\usepackage{braket}
\usepackage{caption}
\usepackage{geometry}
\begin{document}

\title{The non-Hermitian geometrical property of 1D Lieb lattice under Majorana's stellar representation}

\author{Xingran Xu$^{1,2,3}$, Haodi Liu$^{4}$, Zhidong Zhang$^{1,2}$ and Zhaoxin Liang$^{3\dagger}$}

\address{
$^{1}$Shenyang National Laboratory for Materials Science, Institute of Metal Research, Chinese Academy of Sciences, Shenyang, 110016, China\\
$^{2}$School of Materials Science and Engineering, University of Science and Technology of China, Hefei, 230026, China\\
$^{3}$Department of Physics, Zhejiang Normal University, Jinhua, 321004, China\\
$^{4}$Center for Quantum Sciences and School of Physics, Northeast Normal University, Changchun 130024, China}
\ead{$^{\dagger}$ zhxliang@gmail.com}
%\vspace{10pt}
\begin{indented}
\item[]April 2020
\end{indented}

\begin{abstract}
The topological properties of non-Hermitian Hamiltonian is a hot topic, and the theoretical studies along this research line are usually based on the two-level non-Hermitian Hamiltonian (or, equivalently, a spin-$1/2$ non-Hermitian Hamiltonian). We are motivated to study the geometrical phases of a three-level Lieb lattice model (or, equivalently, a spin-$1$ non-Hermitian Hamiltonian) with the flat band in the context of a polariton condensate. The topological invariants are calculated by both winding numbers in the Brillouin zone and the geometrical phase of Majorana stars on the Bloch sphere. Besides, we provide an intuitive way to study the topological phase transformation with the higher spin, and the flat band offers a platform to define the topological phase transition on the Bloch sphere. According to the trajectories of the Majorana stars, we calculate the geometrical phases of the Majorana stars. We study the Lieb lattice with a complex hopping and find their phases have a jump when the parameters change from the trivial phase to the topological phase. The correlation phase of Majorana stars will rise along with the increase of the imaginary parts of the hopping energy. Besides, we also study the Lieb lattice with different intracell hopping and calculate the geometrical phases of the model using non-Bloch factor under the Majorana's stellar representation. In this case, the correlation phases will always be zero because of the normalized coefficient is always a purely real number and the phase transition is vividly shown with the geometrical phases of the Majorana stars calculated by the mean values of the total phases of both right and the joint left eigenstates.
\end{abstract}

%\maketitle

\submitto{\JPCM}
%\maketitle
\captionsetup{font={small,sf},labelfont=bf}
\ioptwocol
\section{Introduction}\label{secintro}
By mapping a high-dimensional projective Hilbert space onto the two-dimensional Bloch sphere, Majorana's stellar representation (MSR)~\cite{Majorana1932} provides an intuitive way to investigate the geometrical phase, dynamics, and entanglement of a high-dimensional spin system~\cite{Ribeiro2007,Ribeiro2008,Bruno2012,Liu2014,Liu2016,Tamate2011,UshaDevi2012,Aulbach2010,Martin2010,Markham2011,Bastin2009,Ribeiro2011,Ganczarek2012,Baguette2014}. In more detail, one can use one point on a $2J+1$-dimensional geometric structure to describe a spin-$J$ state. Alternatively, at the heart of MSR is to describe a
spin-$J$ state intuitively by 2$J$ points on the two-dimensional
Bloch sphere, and these 2$J$ points are
called Majorana stars of the system. At present, there are significant interests and ongoing efforts in investigating such kind of geometrical representation, e.g., the studying of spin-orbit coupling in cold atom physics with the large-spin atoms~\cite{StamperKurn2013,Lian2012,Cui2012,Barnett2009,Lamacraft2010,Whittaker2018}.

Along with the increase of the quality of laser and microcavity technology, a more massive amount of condensate models are realized in the optical and photonic system, which provides a novel platform for achieving the large-spin quantum systems. By the-state-of-art artificial lattice, not only the Hermitian Hamiltonian but also the non-Hermitian Hamiltonian with complex hopping can be created. One of the most attractive models is the Lieb lattice with flat band dispersion~\cite{Vicencio2015}. The localized states are found in Lieb lattice without disorders or defects, and the parity-time($\mathcal{PT}$) symmetric band structures can exhibit non-Hermitian degeneracies known as exceptional points~\cite{Leykam2017,Rev2019}. However, the topological properties of the Lieb lattice are more complicated for the flat band, and topological bands are degenerate. Besides, we can consider Hamiltonian of Lieb lattice in momentum space as pseudospin-$1$, and trajectories of MSR can observe the topological phase transition.

At the same time, the non-Hermitian topological phase transition is a hot topic in contemporary condensate physics. The energy band of the non-Hermitian system can be a complex number, and the eigenstates of the Hamiltonian are not orthogonal any more. However, generalized Brillouin zone(GBZ) extends the knowledge of topological transition by replacing the Bloch factor $\exp\left(ik\right)$ with $\beta$. If $\left|\beta\right|\neq1$, all eigenstates are localized at one of the sides of real space which is famous for `skin effect'~\cite{song2019,yao2018}. The wavefunction of non-Hermitian Hamiltonians can be considered the combination of the wavefunction of every site. However, the spin-$1$ system has three eigenstates, and not all eigenstates have the topological transition. In this case, the skin effect cannot be found by the combination of the system's wavefunction according to the GBZ theory. Besides, the parity symmetry and the parity-time symmetry will make the skin effect disappear, and this phenomenon has been reported in both experiments and theories~\cite{Kawabata2019,Esaki2011,Parto2018,Lieu2018,Weimann2016}. The topological invariants for one dimension system are winding numbers, which is an integer when the system in nontrivial topological phases. In the spin- 1/2 system, the winding number can be characterized as loops, which accounts for times of the trajectories passing around the z-axis connecting north and south poles \cite{jiang2018,Lieu2018} on the Bloch sphere. For the spin-1 system, the zero modes are not two poles points, but the MSR of the flat band provides a tool to define the winding number even on the higher spin. 

In this work, we introduce a three-level toy model which can be realized in the recent photonic or exciton-polariton Lieb lattice with the complex hopping. The topological invariants are surrounded by the generalized winding number, and the edge states can be found in the topological phase in both Hermitian and non-Hermitian regions. Every state of the spin-$1$ system can be represented by two stars on the Bloch sphere, according to the theory of MSR. So the geometrical phases of each state can be divided into two independent phases and one relative phase. We will use MSR to calculate each phase of the system and give the critical point of the phase transition. In the non-Hermitian system, the right and the left eigenstates are both significant. Therefore, we will use the mean values of the total phases to define the topological phase transitions. The Hamiltonian with and without skin effect are both considered in this work, and the numerical and the analytical methods are both included.

The paper is organized as follows. In Sec.~\ref{secmodel}, we introduce our theoretical toy model according to the recent exciton-polariton experimental condition. In Sec. \ref{secMS1}, we give a brief introduction of the Majorana's stellar representation. In Sec.~\ref{sectopo}, we present the method to classify the topological phase in the non-Hermitian system with edge states and the generalized winding number. Then, in Sec.~\ref{secMSR} the trajectories and the geometric phases of Majorana stars are studied in detail to illustrate our theory. Moreover, we consider the system has different intracell hopping give the geometrical phases of the Majorana stars in Sec. \ref{skinef}. A brief discussion and summary are given in Sec.~\ref{secdis}.

\section{Model}\label{secmodel}

The Lieb lattice with the flat band is realized in several systems, including optical, photonic lattices, graphene superconductors, and polariton condensates~\cite{Peotta2015,Mukherjee2015,Qi2018,Ge2018,Chen_2019}. Among these systems, exciton-polaritons are ideal low dimension topological materials which can be realized in recent experiments. Exciton-polaritons are quasiparticles that have a strong coupling between excitons and photons, and they can be applied to realized room temperature Bose-Einstein condensates for its ultra-light effective mass~\cite{Rev0,Rev1,Rev2,Rev3}. Recently, the spin-orbit coupling is applied in exciton-polariton condensates\cite{Whittaker2018}, where the coupling is the pseudospin of polaritons and the photonic orbitals. The non-Hermitian Lieb lattice in the exciton-polariton system can be designed with gain and loss with the technology of buried mesa traps~\cite{Baboux2016,Whittaker2018,Gao2018}.

\begin{figure}
\centering
\includegraphics[width=0.5\textwidth]{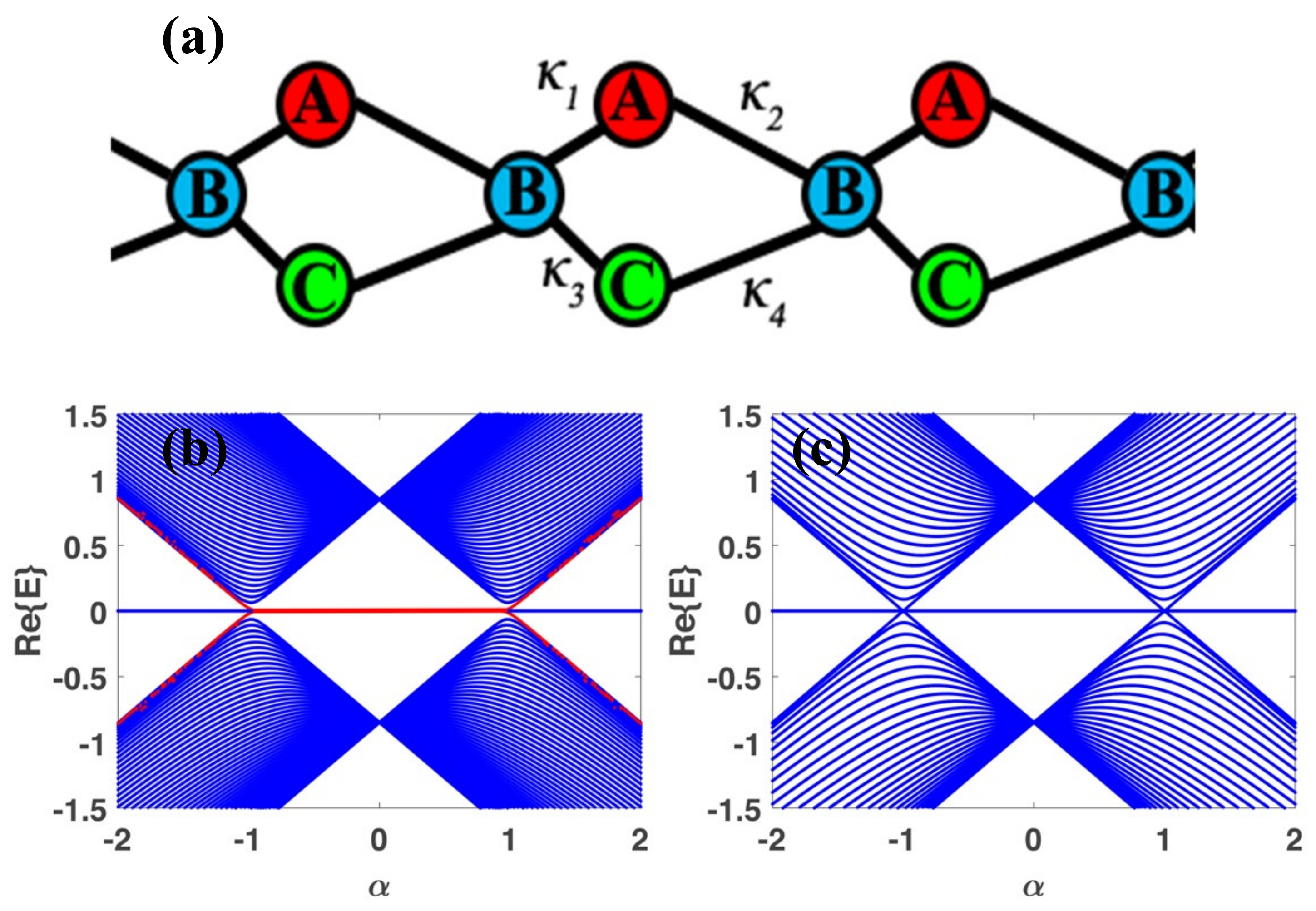}\\
\caption{(a) Illustration of a non-Hermitian Lieb lattice with complex coupling $\kappa_j$, (b), and (c) the energy band of the Lieb lattice with $\lambda$=1 and $\gamma$=0.8 with open boundary condition and periodic boundary condition. }\label{band}
\end{figure}
In this section, we will consider a Lieb lattice with complex hopping in exciton-polariton condensates. As is shown in Fig.~\ref{band}(a), the lattice has three types of site in unit cell: $A$, $B$, and $C$ by setting different pillars in the quantum well. The $A$ and $C$ sites are polaritons with different polarization( parallel and perpendicular)\cite{Xu2017}, meanwhile, $B_n$ sites prevents propagation of flat-band states. This artificial lattice has three dispersive bands, where $A$ and $C$ have the same energy and $B$ exhibit flat band~\cite{Vidal2000,Benoit2002}. In the following sections, we want consider the system in two conditions: one doesn't have skin effect and the other one has skin effect. The MSR will be applied in our calculation to classify and define the topological transitions.
We will focus on the system without skin effect first to study how the geometrical phases will change along with the complex hooping. By using tight-binding approximation, the Hamiltonian of 1D Lieb lattice can be described as:
\begin{eqnarray}
H&=&\kappa_1 \left(a_n^{\dagger}b_{n}+h.c.\right)+ \kappa_2 \left(a_n^{\dagger}b_{n+1}+h.c.\right)\nonumber\\
&+&\kappa_3 \left(c_n^{\dagger}b_n+h.c.\right)+\kappa_4 \left(c_n^{\dagger}b_{n+1}+h.c.\right),
\end{eqnarray}
here, the hopping energy between each site will be a complex number for the gain and loss of the cavity. We will consider $\kappa_2=\lambda+\gamma i$, $\kappa_1=\alpha\kappa_2$, $\kappa_3=\kappa_1^*$, and $\kappa_4=\kappa_2^*$ which ensures the Hamiltonian is $\mathcal{PT}$-symmetric.

The real space energy band of Hamiltonian is illustrated in Fig.~\ref{band}(b) with open boundary condition, where the flat band always appears in the middle of the energy band, and the gap between the 1st and the 2nd (3rd and the 2nd) band will close and open again along with the change of $\alpha $. The flat band and zero-mode of the other two bands are degeneracy, which brings an overwhelmingly strange phenomenon in this system. The $\mathcal{PT}$-symmetry breaking transition and exceptional points occur at $\gamma=\lambda$, and the energy band will change from purely real to the purely imaginary.

Using Fourier transformation, we can rewrite Hamiltonian in momentum space~\cite{Leykam2017}:
\begin{eqnarray}
H=\left(\begin{array}{ccc}
0 & \kappa_{1}+\kappa_{2}e^{ik} & 0\\
\kappa_{1}+\kappa_{2}e^{-ik} & 0 & \kappa_{3}+\kappa_{4}e^{-ik}\\
0 & \kappa_{3}+\kappa_{4}e^{ik} & 0
\end{array}\right),\label{Hmodel}
\end{eqnarray}
with $k$ is the momentum in the Brillouin zone. The first and the third bands are symmetric about the second band(flat band), for the trace of the Hamiltonian is zero. The eigenvalues of Eq.~(\ref{Hmodel}) are
\begin{equation}
E=0,\mp\sqrt{8\left(\lambda^2-\gamma^2\right)\left( 1+\alpha^2+2\alpha\cos k\right)},\label{Ek}
\end{equation}
where the Dirac points will appear at $\alpha=\pm1$. If we just take the periodic condition, there are no zero modes in the first and the third bands, as is shown in Fig.~\ref{band}(c). The zero-mode in the first or the third energy band is the energy of the boundary states. The right eigenvectors of the system are :
\begin{eqnarray}
\psi_{1,R}=\left(\begin{array}{c}
\frac{2\gamma}{\gamma-i\lambda}-1\\
0\\
1
\end{array}\right),
\psi_{2(3),R}=\left(\begin{array}{c}
1-\frac{2\gamma}{\gamma+i\lambda}\\
\frac{iE_{2(3)}}{\left(\gamma+i\lambda\right)\left(\alpha+e^{ik}\right)}\\
1
\end{array}\right),\label{eigenstates}
\end{eqnarray}
here, eigenstates $\psi_{2(3),R}$ are not normalized or orthogonal, which is the characteristic property of the non-Hermitian system. $\psi_{1,R}$ is the eigenstates of the flat band, and the density of $B$ sites is always zero. Besides, for the other two bands only when $E_{2(3)}$ is zero, the density of $B$ sites is zero. If zero modes appear with a flat band in this system, there are no condensates in $B$ sites. We will use the left eigenstates and the right eigenstates of the Hamiltonian together to define the topological invariants and the orthogonality.

\section{ Majorana's stellar representation }\label{secMS1}
For this spin-$1$ system, one convenient and intuitive way to study symmetry and dynamics of its quantum state is on the Bloch sphere. For a two-level system, the pure state can be naturally mapped on the Bloch sphere with $\xi=\tan\theta e^{i\phi}$. The momentum can change in a Brillouin zone which drives the point $\vec{u}=(\theta,\phi)$ moving periodically. Therefore, one can visualize the Berry phase for the spin $1/2$ state by the solid angle subtended by the close trajectory of $u$ on the Bloch sphere.
However, this geometric interpretation is hard to be represented intuitively in a higher dimensional Hilbert space, even we can map the quantum pure state to a complex higher-dimensional geometric structure. By using MSR, the Berry phase, Berry connection, and Berry curvature, which are related to the topological structure of the spin-$J$ system, can be intuitively studied by $2J$ stars on a two-dimensional Bloch sphere\cite{Liu2014,Liu2016}. The geometrical phase of a large spin system then can be studied by the trajectories of Majorana stars. These Majorana stars can be derived by parameterizing quantum state $\psi=\sum C_m\ket{m}$ as a series of complex number $\zeta^{\left(m\right)}=\tan\frac{\theta_m}{2}e^{i\phi_m}$ with
\begin{equation}
\sum_{m=0}^{2n}\left(-1\right)^{m}\frac{C_{n-m}}{\sqrt{\left(2j-m\right)!m!}}\zeta^{\left(m\right)}=0.
\end{equation}
The state $\psi$ can then be mapped onto the Bloch sphere as stars $\vec{u}_m=(\theta_m,\phi_m)$.
In the MSR, the Berry phases for the eigenstates of the spin-$1$ system can also be described intuitively as\cite{Liu2016}
\begin{eqnarray}
\gamma_b^{\left(n\right)}&=&\gamma_0^{\left(1\right)}+\gamma_0^{\left(2\right)}+\gamma_C\\
&=&-\int \frac{1-\cos\theta_1 }{2}d\phi_1-\int \frac{1-\cos\theta_2 }{2}d\phi_2\nonumber \\
&+&\frac{1}{2}\oint\frac{\left(d\vec{u}_{1}-d\vec{u}_{2}\right)\cdot\left(\vec{u}_{1}\times\vec{u}_{2}\right)}{3+\vec{u}_{1}\cdot\vec{u}_{2}}, \label{HMphase}
\end{eqnarray}
where $u_1=(\theta_1,\phi_1)$ and $u_2=(\theta_2,\phi_2)$ are the coordinates of Majorana stars, $\gamma_0^{\left(1,2\right)}$ are two independent solid angles on the Bloch sphere and $\gamma_C$ is the correlation phase between two Majorana stars. Except the independent solid angles $\gamma_0^{1,2}$ subtended by the closed evolution paths of the Majorana stars, the geometric phase also contains the correlation phase $\gamma_C$ brought by the correlation between the stars \cite{Liu2014,Liu2016}. In this paper, we will build the relation of the non-Hermitian system's geometrical phases and the MSR. The correlation phase is also corresponding to the imaginary parts of the normalized coefficient of the system.

The Berry phase of the non-Hermitian system can be defined by using quantum geometric tensor \cite{Zhang2019}. However, we want to use  both right and the joint right eigenstates to define the geometrical phases. For a non-Hermitian system $H\neq H^{\dagger}$, the left and the joint right eigenstates are generally unrelated, although they share the same eigenvalues\cite{Shen2018}. The geometrical phases of the Majorana stars of the non-Hermitian Hamiltonian also need to consider the right and the joint left eigenstates. The winding number defined by eigenstates is equal to the summation of the geometrical phases of these eigenstates \cite{jiang2018}.

\section{Edge states and Non-Hermitian winding number of the topological Lieb lattice}\label{sectopo}
In this section, we will solve the edge states and discuss the topological invariants of the non-Hermitian system. The bulk-edge correspondence in Hermitian and the non-Hermitian system is very different. Recently, skin effect is found in the non-Hermitian system, where all bulk states are localized at one side in real space. Besides, the skin effect is relative to the generalized Bloch zone, and the Bloch factor $\left|e^{ik}\right|$ can be smaller or bigger than 1. If the Bloch factor is not equal to 1, the skin effect can be found in this system. Otherwise, the skin effect will not exist, and the Bloch theory still works.

\begin{figure}
\centering
\includegraphics[width=0.5\textwidth]{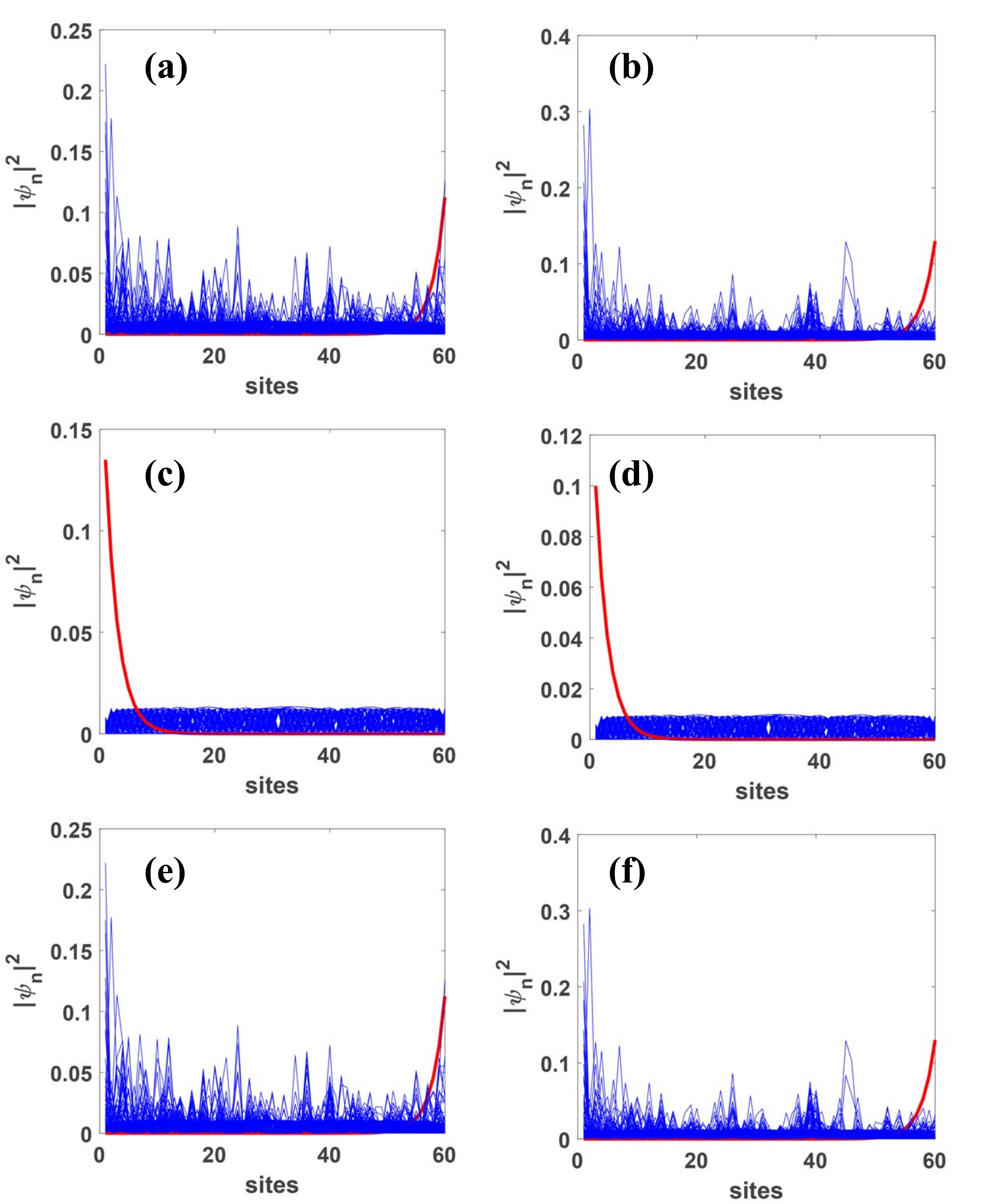}\\
\caption{Eigenstate wave functions on sublattice A (the first row), site B (the second row), and site C (the third row. We take $\alpha$=0.8, $\lambda$=1, $\gamma$=0.5 for the first column, and $\gamma$=1.5 for the second column. }\label{wf}
\end{figure}

The energy of edge states are shown in Fig.~\ref{band}(b) with red lines where zero modes will appear according to the open boundary condition. The Eigen equations of the first unit in real space can be described as:
\begin{eqnarray}
E\psi_{1,A}&=&\kappa_1 \psi_{1,B}+\kappa_2 \psi_{2,B},\label{edge1}\\
E\psi_{1,B}&=&\kappa_1 \psi_{1,A}+\kappa_3 \psi_{1,C},\label{edge2}\\
E\psi_{1,C}&=&\kappa_3 \psi_{1,B}+\kappa_4 \psi_{2,B}.\label{edge3}
\end{eqnarray}
we can substitute Eqs. (\ref{edge1})-(\ref{edge2}) to Eq. (\ref{edge3}) and get the relation between $\psi_{1,B}$ and $\psi_{2,B}$ and get:
\begin{eqnarray}
\psi_{2,B}&=&-\frac{E^2-\kappa_1^2-\kappa_2^2}{\kappa_1\kappa_2+\kappa_3\kappa_4}\psi_{1,B}\nonumber\\
&=&-\left(\alpha+\frac{E^2}{\kappa_1\kappa_2+\kappa_3\kappa_4}\right)\psi_{1,B}.
\end{eqnarray}
From Fig.~\ref{band}(b), the zero-mode can appear at the region of $\left|\alpha\right|<1$ and the edge density of sites B will have a localized rate of $\left|\alpha\right|$ as is vividly shown in Figs.~\ref{wf}(c) and (d). Besides, the relation between the density distribution of $A$ and $C$ sites can be found from $\psi_{1,C}=\frac{E\psi_{1,B}-\kappa_1\psi_{1,A}}{\kappa_3}$ which reveals the density distribution of these two sites are the same if the system is in the topological nontrivial phase region. If the system is in the topologically trivial phase, the density distribution of $A$ and $C$ sites are very different.

To define the winding number and discuss the orthogonality of the system, the joint left eigenstates need to be calculated and combine them with the right eigenstates. The joint left eigenstates are satisfied $H^{\dagger}\ket{\psi_{n,L}}=E^*\ket{\psi_{n,L}}$ comparing to $H\ket{\psi_{n,R}}=E\ket{\psi_{n,R}}$. The eigenvectors and t the joint eigenvectors are biorthogonal and complete:
\begin{eqnarray}
&\braket{\psi_{m,L}|\psi_{n,R}}=0 \left(m\neq n\right),\\
&\sum_n\frac{\ket{\psi_{n,R}}\bra{\psi_{n,L} }}{\braket{\psi_{n,L}|\psi_{n,R}}}=\hat{1},\label{guiyi}
\end{eqnarray}
here, the normalization convention is enforced but not necessary. The geometric phase for a biorthogonal system is given by \cite{Hayward2018}
\begin{equation}
\gamma_b=\int_{0}^{z}\frac{\bra{\psi_{n,L}}i\partial_{z'}\ket{\psi_{n,R}} }{\braket{\psi_{n,L}|\psi_{n,R}}}dz',\label{biphase}
\end{equation}
with $z'$ is the dependent parameters. It is different from the usual Berry phase that the normalization of the system can be a complex number. In our model, if dissipative term $\gamma$ is beyond zero, the Hamiltonian is non-Hermitian, and the bi-normalized number of eigenstates will be complex. The energy will change from purely real to purely imaginary, along with an increase of $\gamma$.

There is no skin effect in this non-Hermitian system because of the $\mathcal{PT}$-symmetry. The winding number of the non-Hermitian system can be described in Bloch zone as
\begin{eqnarray}
W_n&=&\frac{1}{\pi}\int_{-\pi}^{\pi}\frac{ \bra{\psi_{n,L}}i\partial_k\ket{\psi_{n,R}} }{\braket{\psi_{n,L}|\psi_{n,R}}}dk,\\
&=&\left\{ \begin{array}{c}
\pm1,\left|\alpha\right|<1,\label{wingdingDF}\\
0,\left|\alpha\right|\geq1.
\end{array}\right.\label{winding}
\end{eqnarray}
for $n=2,3$. From Eq.~(\ref{winding}), we can find $\alpha\in\left(-1,1\right)$ the winding number is 1, otherwise the winding number is zero which can be described as the topological invariant.

As is shown in Fig.~\ref{wf}, the edge states are plotted with red lines in their real space wavefunction. Edge states of $A$ sites and $C$ sites are localized at the last unit cell of the system. Relatively, $B$ sites are localized at the first unit cell of the system. The parameters of the first column of Fig.~\ref{wf} are $\mathcal{PT}$-symmetry unbroken, and the second column is symmetry broken. Along with the increase of the gain and loss, edge states always exist, and the skin effect does not appear in the topological phase region.

\section{The Majorana Representation and geometric phases on the Bloch sphere }\label{secMSR}

\begin{figure}
\centering
\includegraphics[width=0.5\textwidth]{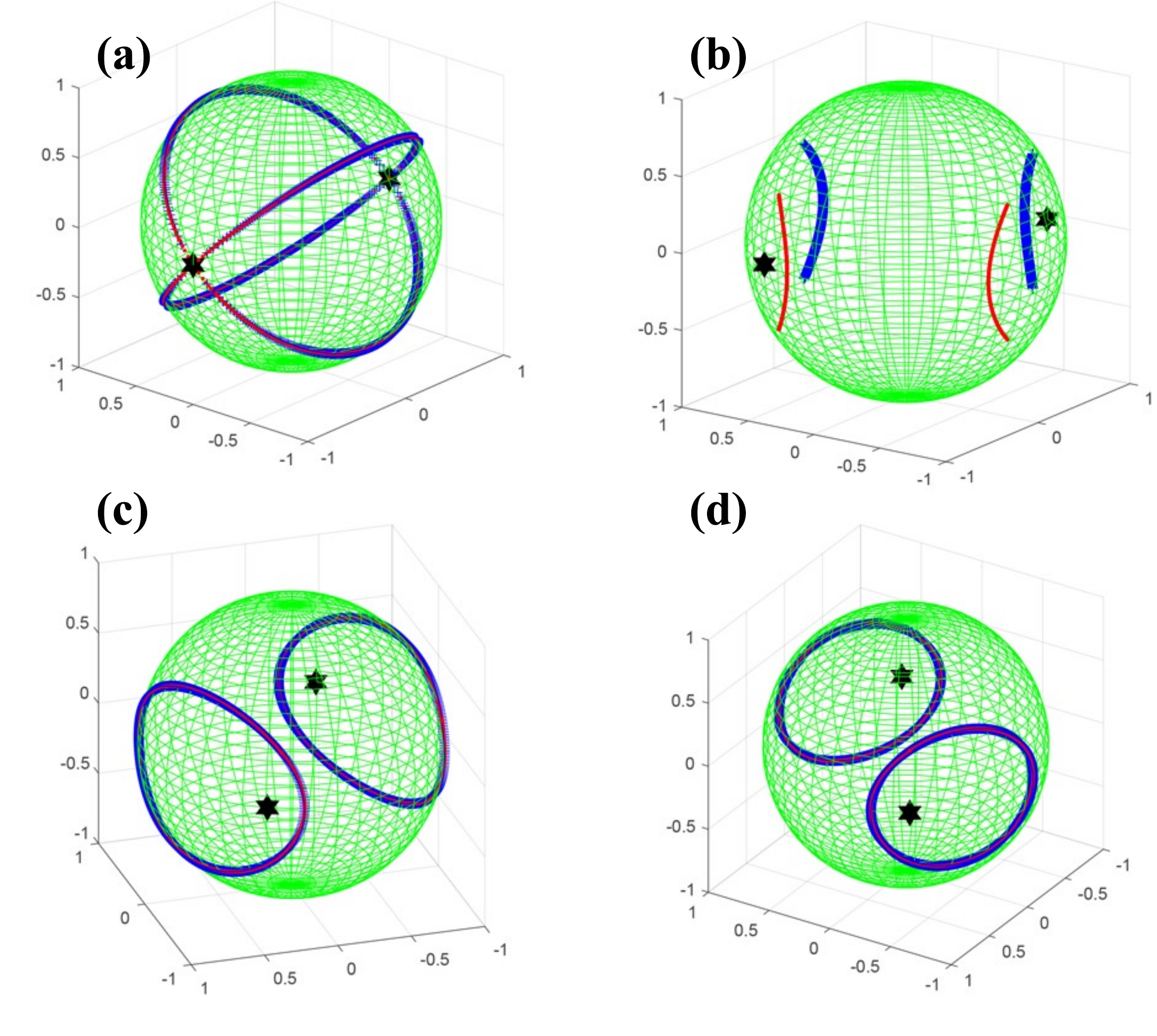}\\
\caption{Majorana stars of the eigenvectors on Bloch sphere. The dark stars are MSR of $\zeta^{\left(1\right)}$, red and blue lines are trajectories of MSRs of $\zeta^{\left(2\right)}$ and $\zeta^{\left(3\right)}$. Parameters are used: $\lambda$=1, and (a) $\alpha$=0.0. $\gamma$=0.5; (b) $\alpha$=2, $\gamma$=0.5; (c) $\alpha$=0.5, $\gamma$=0.5; and (d) $\alpha$=0.5, $\gamma$=1.5. }\label{MSR}
\end{figure}
It is worthy to notice that our system can be interpreted by a pseudospin-$1$ model. Comparing to the spin-$1/2$ system, the geometrical phases of a large spin system are more complicated because the correlation phases need to be considered.

The Majorana stars of flat band's eigenvector on Bloch sphere are two fixed points with $\zeta^{\left(1\right)}=\pm \sqrt{-1+\frac{2\gamma}{\gamma-i\lambda}}$ and these two points can be considered as the singular points. If $\gamma$ is zero, the points of MSR of $\zeta^{\left(1\right)}$ are $\left[\pm1,0,0\right]$ which is the same with $x$ axis for spin-$1$ system. Majorana stars of the rest two eigenstates can be represented on Bloch sphere with
\begin{eqnarray}
\zeta^{(2,3)}&=&\frac{iE_{2,3}}{\sqrt{2}(\gamma+i\lambda)\left(1+\alpha e^{ik}\right)}\nonumber\\
&\pm& i\sqrt{\frac{\gamma+i\lambda}{\gamma-i\lambda}+\frac{E_{2,3}^{2}}{2(\gamma+i\lambda)^{2}\left(1+\alpha e^{ik}\right)^{2}}},
\end{eqnarray}
here, if $E_{2,3}$ are zero, $\zeta^{\left(2,3\right)}$ have the same fixed points of $\zeta^{\left(1\right)}$. If $E_{2,3}$ are not zero, the Majorana stars will move on the Bloch sphere along with the change of the momentum.
\begin{figure*}
\centering
\includegraphics[width=0.9\textwidth]{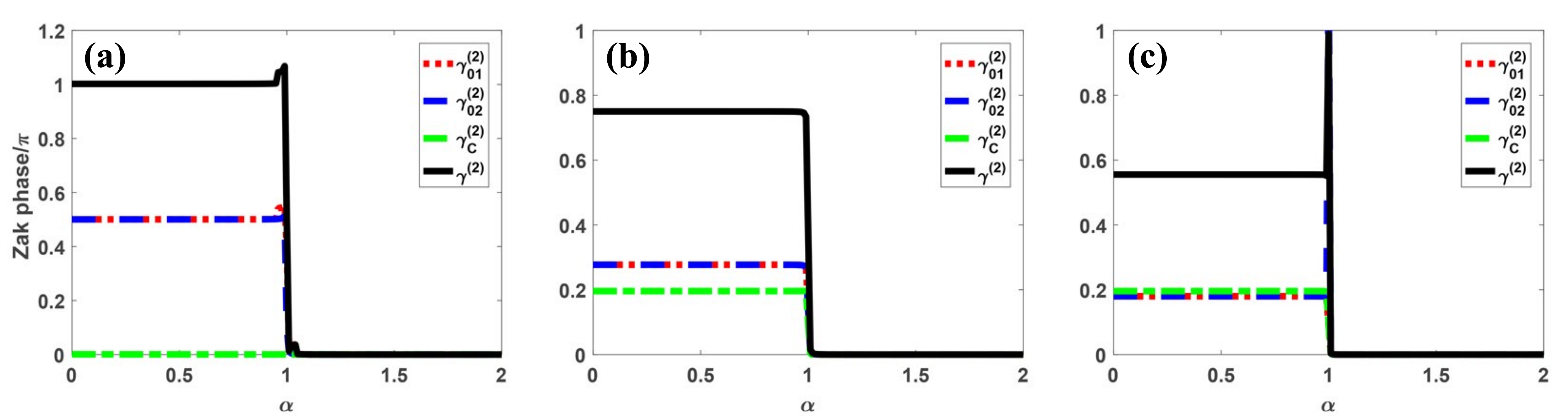}\\
\caption{Geometrical phases of trajectories of Majorana stars $\zeta^{\left(2\right)}$ on Bloch sphere using Eq.~(\ref{HMphase}). We take $\lambda$=1, $\alpha$=0.5, $\gamma$=0, 0.5,1.5 for (a), (b), and (c). }\label{ZP}
\end{figure*}

Although, we present the usual way to calculate the winding number in Sec.~\ref{sectopo}. MSR provides another way to calculate the winding number. For the spin-$1/2$ Hermitian system we can just use $\nu=\frac{1}{2\pi}\int \frac{\partial \phi\left(k\right)}{\partial k} dk$, which is the loops of the trajectory of the state on the Bloch sphere goes around $z$ axis\cite{jiang2018}. But it is not useful for non-Hermitian system with the complex normalized coefficient. In our model, if $\gamma$ is zero, Majorana of stars of $\psi_1$ are $[-1,0,0]$ and $[1,0,0]$ and their connection line can be considered as $x$ axis. The winding number can be defined by $\nu_{tol}^{\left(n\right)}=\nu_1^{\left(n\right)}+\nu_2^{\left(n\right)}$ with $\nu_{1,2}^{\left(n\right)}$ are the winding number of the Majorana stars go around the every singularity points of $\psi_n$ state.

The normalized coefficients of each state are $\frac{2 (\gamma^2 -\lambda^2 ) }{(\gamma -i \lambda )^2}$, $\frac{4 (\gamma^2 -\lambda^2 ) }{(\gamma +i \lambda )^2}$, and $\frac{4 (\gamma^2 -\lambda^2 ) }{(\gamma +i \lambda )^2}$ by using Eq.(\ref{guiyi}). The normalized coefficients will change with dissipative term $\gamma$ which makes the phase of the Bloch factor change from $e^{ik}$ to $e^{ik+i\phi}$ with $\phi$ is an additional phase given by the angles of the normalized coefficients. The trajectories Majorana stars of eigenstates will also change with $\gamma$ because the additional phase $\phi$ makes an angle shift of the Bloch sphere.

Since the Hamiltonian has $\mathcal{PT}$-symmetry broken and unbroken phases in different parameters, we can first focus on the Hermitian condition by setting $\gamma$ to zero. As shown in Fig. \ref{MSR}, the dark state ($\psi_1$) forms two fixed stars on the Bloch sphere and can not move along with different momentum. While stars of $\psi_{2,R}$ and $\psi_{3,R}$ can move along with the changing of different momentum, and their trajectories will cross at the fixed points of $\psi_{1,R}$. The trajectories of $\psi_{2,R}$ and $\psi_{3,R}$ will not come across and form a complete circle when $\alpha$ is larger than $1$. However, as $\alpha$ decreases, they will get close to each other. Besides, only $\alpha<1$, $\psi_{2,R}$ and $\psi_{3,R}$'s trajectories can form a closed loop, and they will coincide as shown in Figs \ref{MSR}(c) and (d). The Berry phase can be calculated by using Eq. (\ref{HMphase}) as the solid angles of the Majorana stars in the Bloch sphere, which is the same result of the calculation of the Zak phase.

As is illustrated in Fig.~\ref{ZP}, if $\gamma$=0, the geometric phase on the Bloch sphere has a jump at $\alpha$=1, and the total phase arises from zero to $\pi$. The correlation phase is always zero, which means that there is no correlation between the two stars. The contributions of the geometric phases from the two stars' independent evolution are the same and equal to $\pi/2$. When the non-Hermitian part $\gamma$ is introduced, the total phase decreases, and the correlation between the two stars arise. As $\gamma$ increases, the correlation phase increase, well, the total phase and its noncorrelation part decrease. This means that the correlation between the stars can significantly influence the geometric phase and also can be related to the degree of the non-Hermitian.

Remarkably, we only calculate the geometric phases of all right eigenstates. However, the left eigenstates are already included. $\psi_{1,R}$ is always topological trivial, and it is the eigenstates of the flat band, so the joint left eigenstates of $\psi_{1,R}$ is itself. Due to the $\mathcal{PT}$ symmetry, we can easily find the joint left eigenvector of $\psi_{2,R}$ is $\psi_{3,R}$ and $\psi_{2,R}$ is the joint left eigenstates of $\psi_{3,R}$. Like the Majorana stars shown in Fig.~\ref{MSR}(b), the Majorana stars of left and the right eigenstates for the same eigenvalues are around the same singular points with different colors. The geometrical phase we show in Fig.~\ref{ZP} is different from Eq.~(\ref{biphase}) because the definition of Majorana stars can only use one side of the eigenstates.

\section{The geometrical phase of the Lieb lattice with skin effects}\label{skinef}

The non-Hermitian topological transition will breakdown the Bloch theory which makes the Bloch factor bigger or smaller than 1. Recently, the skin effect has been observed in the non-Hermitian SSH model both experimentally and theoretically and the geometrical phases of the non-Hermitian system become more and more significant. In this section, we will consider a spin-1 Lieb lattice with  different intracell hopping and use the MSR to study the geometrical phases of the system.

We change the hopping energy $\kappa_2=\kappa_4=t_2$ of Eq. (\ref{Hmodel}) and the Hamiltonian in momentum space can be rewritten as :
\begin{equation}
H=\left(
\begin{array}{ccc}
0 & \kappa_1+ t_2 e^{ik} & 0 \\
\tilde{\kappa}_1+t_2 e^{-ik} & 0 & \tilde{\kappa}_1+t_2 e^{-ik} \\
0 & \kappa_1+ t_2 e^{ik} & 0 \\
\end{array}
\right),\label{skModel}
\end{equation}
here, $\kappa_1=t_1+\delta$ and $\tilde{\kappa}_1=t_1-\delta$ are the hopping energy of different sites and $\delta$ is the intracell hopping difference which can control the non-Hermicity of the system. The trace of Matrix (\ref{skModel}) is zero, so the eigenvalues are $(0,-E, E)$. By diagonalization, the eiegnenergies of the system are
\begin{equation}
E(k)=0, \mp\sqrt{ 2\left[t_2 e^{-ik}+(t_1-\delta) \right]\left(t_1+t_2e^{ik}+\delta \right)},
\end{equation}
and the eigenstates are
\begin{equation}
\psi_{1,R}=\left(\begin{array}{ccc}
- \frac{1}{\sqrt{2}} \\
0\\
\frac{1}{\sqrt{2}}\\
\end{array}\right), \psi_{2(3),R}=\left(\begin{array}{ccc}
\frac{1}{2} \\
\frac{E_{2(3)}/2}{ t_1+t_2e^{i k}+\delta}\\
\frac{1}{2}\\
\end{array}\right).\label{psisk}
\end{equation}
If $\delta$ is zero, the eigenenergies are all real numbers, however, if $\delta$ is not equal to zero, the system will have skin effects and the generalized Brillouin zone needs to be considered \cite{yao2018,Shen2018,deng2019}.

In \ref{appendix1}, we calculate the motion of the Majorana stars and the geometrical phases of the Majorana stars in the Brillouin zone with the Bloch factor $e^{ik}$. The critical point is given by $t_1=\left|t_2-\delta \right|$, however, the geometrical phase is not well defined, and the critical point is different from the zero modes of the edge state as is shown in \ref{appendix2}. Although we consider the Majorana stars of the left and the right eigenstates, the change of the geometrical phase doesn't correspond with the Ref. \cite{yao2018}. If we want to study the topological transition of the system, the momentum needs to be modified.

The edge state of the non-Hermitian system is different from the bulk state. However, the Brillouin zone can not describe the critical point exactly. In recent research, the generalized Brillouin zone is considered in the non-Hermitian system to deal with the topological transition with skin effect. We need to replace $e^{ik}$ with $\beta$ and the momentum $k=-i ln \beta$. The generalized Bloch factor we calculated in the \ref{appendix2}, and we will use this result to study how the non-Bloch factor affects the geometrical phases of Majorana's stars.

The non-Bloch factor of this system is $\beta=\sqrt{ \left|\frac{t_1+\delta}{t_1-\delta}\right| }$, which means we need to modify our above result and use the generalized Bloch factor to classify the topological transition of the system. The critical point of the topological transition is $t_1=\left|t_2-\delta\right|$ in Brillouin zone with real momentum. However, this result doesn't match with the edge mode of the Hamiltonian in the real space. If we use the generalized Bloch factor $\beta$, the energy of the Hamiltonian can be a purely real number. By replacing $e^{ik}$ with $\beta e^{ik}$ and $e^{-ik}$ with $1/\beta e^{-ik}$ in Eq. \ref{Windingsk}, we can get the system's winding number:
\begin{eqnarray}
W_n&=&\frac{1}{2\pi}\oint_{-\pi}^\pi \left(\frac{t_{2}}{t_{2}+\frac{e^{ik}(t_{1}-\delta)}{\sqrt{\left|-\frac{t_{2}-\delta}{t_{2}+\delta}\right|}}}+\frac{\frac{e^{ik}t_{2}}{\sqrt{\left|-\frac{t_{2}-\delta}{t_{2}+\delta}\right|}}}{\frac{e^{ik}t_{2}}{\sqrt{\left|-\frac{t_{2}-\delta}{t_{2}+\delta}\right|}}+\delta+t_{1}}\right)dk.
\end{eqnarray}   
 According to the above equation, we can get the critical point of the topological transition is $t_1=\delta+t_2/\beta$ and $t_1=-\delta+t_2\beta$. As the non-Bloch factor we calculated in \ref{appendix1}, the transition point can be   obtained by :
 \begin{eqnarray}
 t_1^C&=&\sqrt{ \left( t_2/\beta+\delta \right)  \left( t_2\beta-\delta \right)  }.\label{crp}
 \end{eqnarray}
To vertify this result, we will use MSR to calculate the geometrical phases of the system.
 
\begin{figure}
\includegraphics[width=0.5\textwidth]{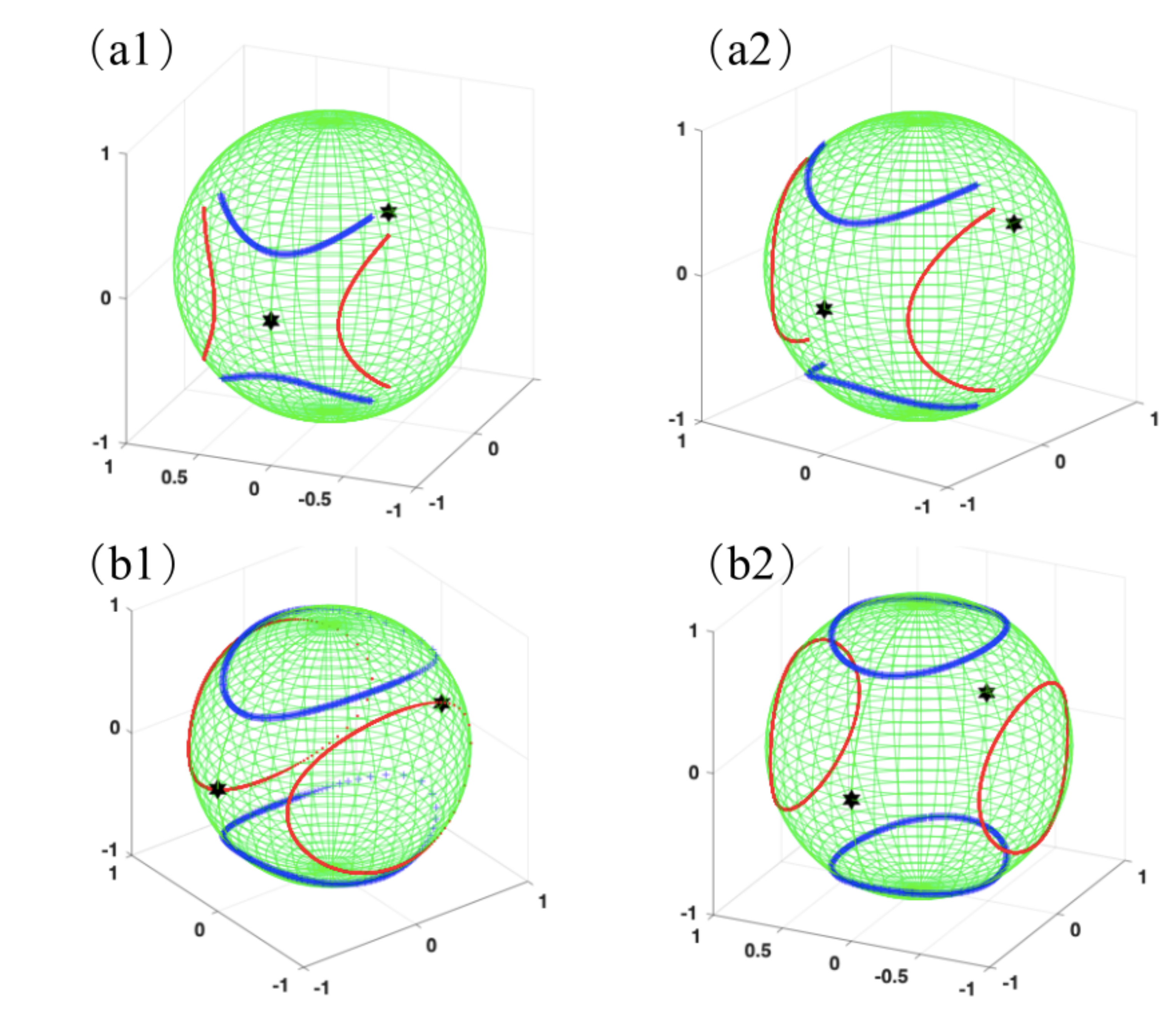}\\
\caption{Majorana stars of the right eigenvectors (red lines) and the joint left eigenvectors (blue lines ) of the $\psi_2$ with generalized Bloch factor $\beta=\sqrt{\left|\frac{t_1+\delta}{t_1-\delta} \right|}$. We take $t_2$=1, $\delta$=0.2, and $t_1$=2, 1.1, 1.01, 0.5 for (a1), (a2), (b1), and (b2). }\label{MSRGBZ}
\end{figure}

As is shown in Fig. \ref{MSRGBZ}, we plot the Majorana stars of $\psi_2$ long with the change of the momentum. When we use generalized Bloch factor, the Majorana stars have a different dynamic behaviour compared to Fig. \ref{skMSR}. In Fig. \ref{MSRGBZ} (a1) and (a2), the Majorana stars move around the zero-mode, but they cannot form a closed loop. The stars of the right eigenstates are moving in the eastern and the western hemisphere, while stars of the left eigenstates are moving in the northern and the southern hemisphere. Therefore, the trajectories of all stars of $\psi_2$ will not intersect. Along with the decrease of $t_1$, the system will change from the trivial phase to the topological phase, and each star can form a closed loop. In particular, the trajectories of the Majorana stars of the left eigenstates can enclose $z$-axis.

\begin{figure}
\includegraphics[width=0.5\textwidth]{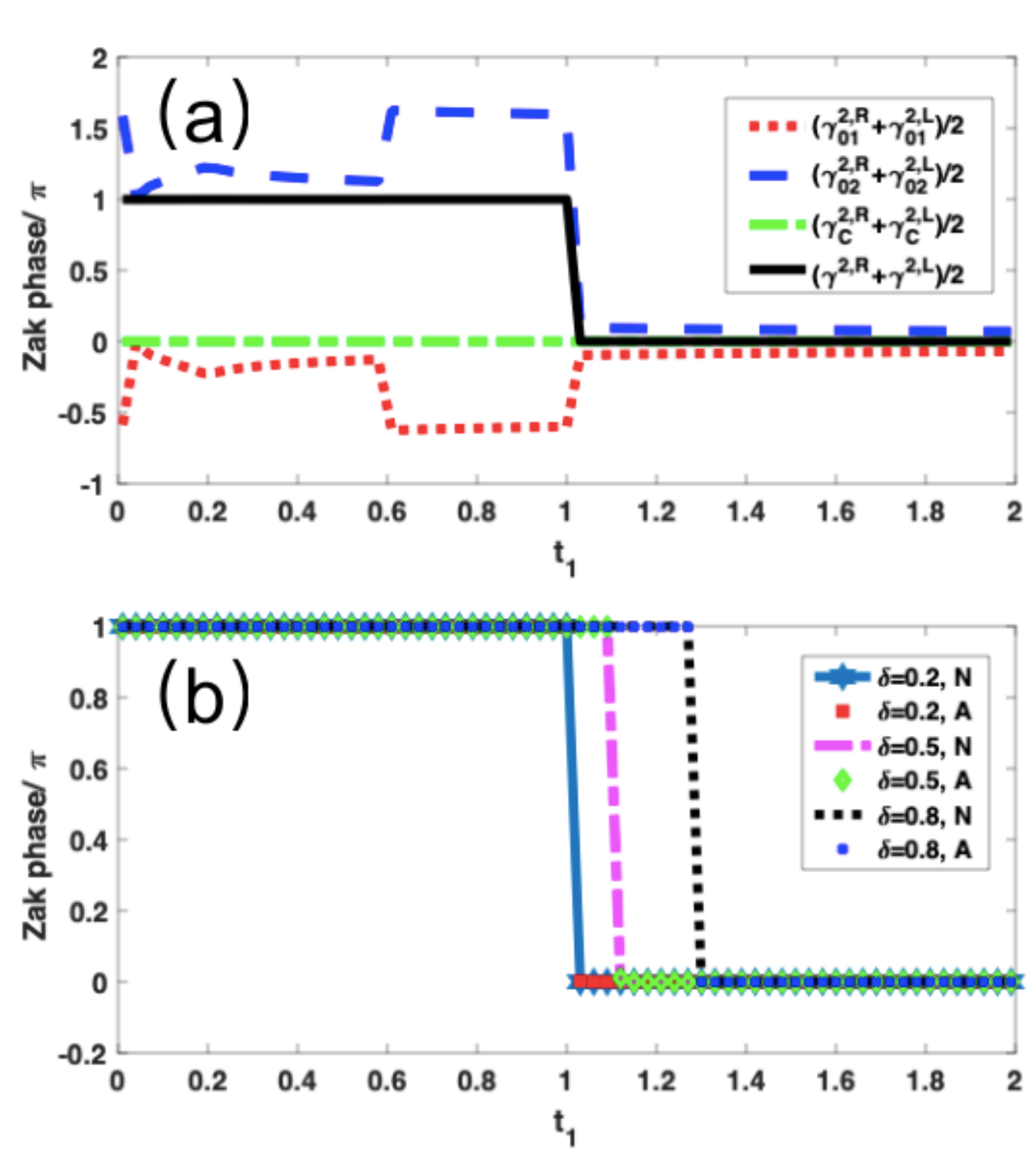}\\
\caption{(a) The mean values of each phase of the right eigenstates and the joint left eigenstates and the total phases of $\psi_2$ with non-Bloch factor $\beta=\sqrt{\left|\frac{t_1+\delta}{t_1-\delta} \right|}$,  $t_2$=1, and $\delta$=0.2. (b) The mean values of the total phases (N) and the analytical  results obtained by the Eq. (\ref{crp}) (A) of the geometrical phases.}\label{MSRGBZ}
\end{figure}

To observe the geometrical phases of the system, we calculate the mean values of each phase with non-Bloch factor in Fig. \ref{MSRGBZ}(a). The mean value of the correlation phase is always zero because the normalized coefficient is a purely real number. The mean value of the total phase have a sudden jump at $t_1^C$, which is different from the result of the Brillouin zone, and the similar result has been obtained by Eq. (\ref{crp}). Although we use a different way to classify the topological transition, the non-Bloch factor still need to be calculated. The geometrical phases defined by MSR and the eigenstate can give the same result.

From above analysis, if we use Majorana's stellar representation to study the non-Hermitian topological transition, we need to take the following steps:
\begin{itemize}
\item Calculate the eigenenergies and the eigenstates of the system.
\item Plot the eigenstates and choose the non-Bloch factor or Bloch factor according to the existence of  the skin effect. 
\item Map the right and the joint left  eigenstates on the Bloch sphere.
\item Calculate the independent phases and the correlation phases of the trajectories of the Majorana stars.
\item The mean values of the total phases of the right and the joint left eigenstates need be considered as the geometrical phases of the system.
\end{itemize}
There are two differences between the Hermitian and the non-Hermitian condition. One is we need to consider the geometrical phases of the right and the joint left eigenstates together, and the other one is the non-Bloch factor need to be considered if the system has skin effect.

\section{Discussion}\label{secdis}
Summarizing, we use Majorana representation to classify the topological transition of two toy models. The Hamiltonian we propose is non-Hermitian with complex hopping or different intracell hopping, which can be realized in recent experiments. Remarkably, we use exciton-polaritons condensates as an example to investigate the non-Hermitian topological phase transition in our main content. Still, this toy model can be realized in other systems like photonic crystal. The bulk-boundary correspondence still exits not only in $\mathcal{PT}$ unbroken but also in the broken region with the localization of the edge states. Besides, the topological invariants can be well defined by winding numbers with both left eigenvectors and right eigenvectors, and the topological region is shown in this paper. We illustrate the trajectories and geometrical property of spin-$1$ Lieb lattice with Majorana presentation. The loops of the trajectories of Majorana stars go around their singular points can be considered as the winding number to classify the topological trivial or nontrivial phase. Moreover, by calculating the berry phase of the MSR, we find the Berry phase of Majorana stars has a jump when the parameters change from trivial phase to the nontrivial phase. Furthermore, the correlation phase arises from zero to nonzero, along with the increase of the dissipative term if the normalized coefficient is a complex number. In contrast, the correlation phase will always be zero if the normalized coefficient is a purely real number. Last but not least, if the system has skin effect, the non-Bloch factor needs to be considered, and the mean values of the total phases of the right the joint left eigenstates under MSR can be the geometrical phases of the system.

{\it Note added.} During the revised stage of our manuscript preparation, we became aware of a preprint \cite{teo2020topological}, which uses MSRs to study the topological transitions, with a very different focus.

\section{Acknowledgements}
We thank W. Yi, T.-S. Deng, and Y. Chen for stimulating discussion. This work is supported by the National Natural Science Foundation of China (Grant No. 11604300, No. 11875103) and Key Projects of the Natural Science Foundation of China (Grant No. 11835011). Z. D. Z. is supported by the NSFC of China (Grant No. 51331006).
\section{References}
\bibliography{mybib}
\appendix

\section{The geometrical phases in the Brillouin zone under the Majorana‘s representation}\label{appendix1}
\begin{figure}
% \centering
\includegraphics[width=0.5\textwidth]{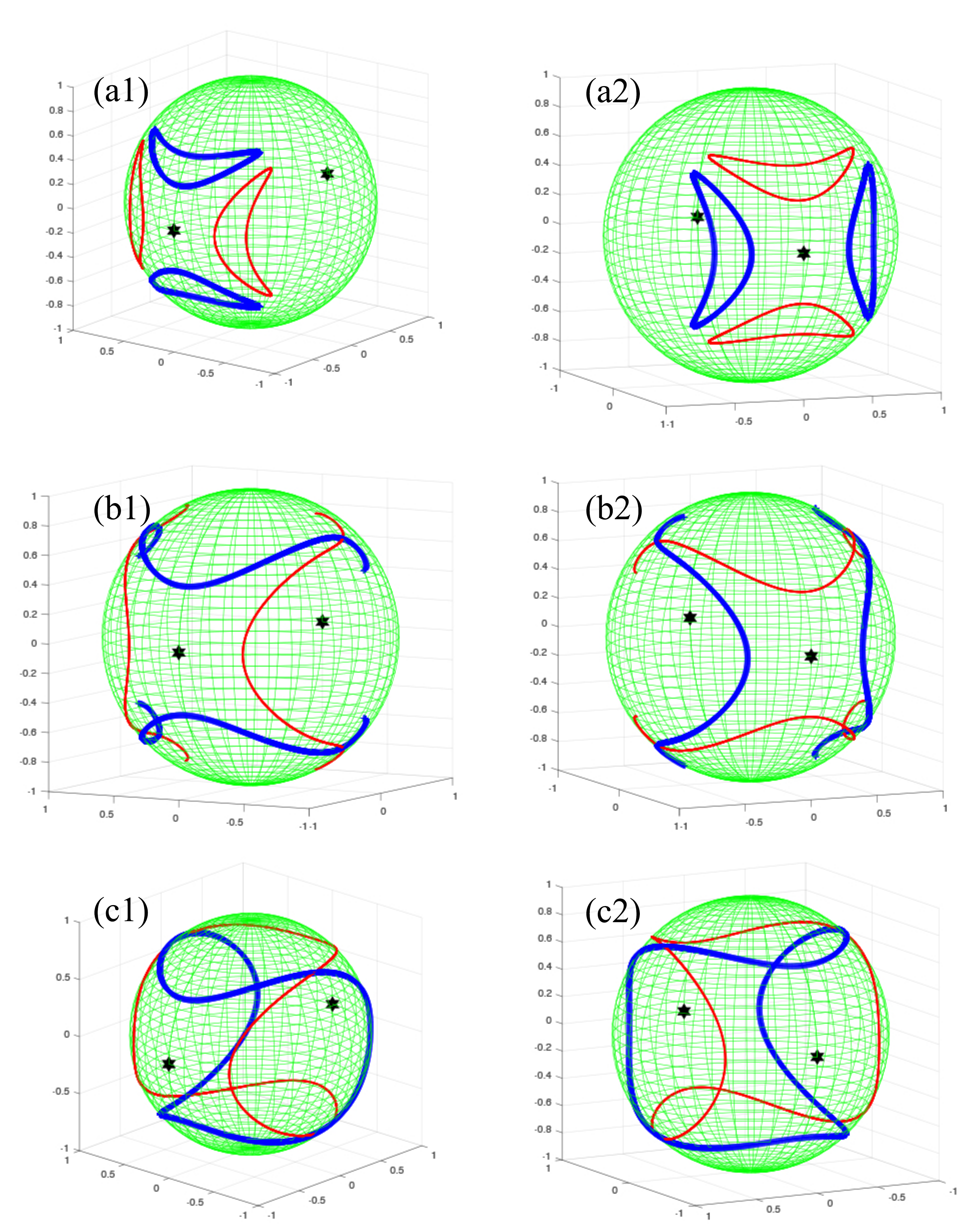}\\
\caption{Majorana stars of the right eigenvectors (red lines) and the joint left eigenvectors (blue lines ) of the $\psi_2$ with skin effect. We take $t_2$=1, $\delta$=0.2, and $t_1$=2, 0.9, 0.2 for the first column, and $t_1$=-2, -0.9, -0.2 for the second column. }\label{skMSR}
\end{figure}

In this appendix, we will calculate the motion of the Majorana stars and the geometrical phases of  the model \ref{skModel}. 
The orthogonality and the normalization can be well defined by Eq. (\ref{guiyi}) and the normalized coefficient is a purely real number. The winding number can be defined by the energy $\nu_E$ as \cite{jiang2018,Leykam_2017,Shen2018,Liu_2020}: $\nu_E=\frac{1}{2\pi}\oint \partial_k \arg(\Delta E )dk$, meanwhile, using Eq. (\ref{wingdingDF}), the left and joint eigenstates can give the winding number as
\begin{eqnarray}
W_n&=&\oint_{-\pi}^\pi \frac{t_2 e^{ik} (-i \delta \sin (k)+t_1\cos (k)+t_2)}{ \left(t_2+e^{i k} (t_1-\delta )\right) \left(\delta +e^{i k} t_2+t_1 \right)} dk,\label{Windingsk}\\
&=&\frac{1}{2\pi}\left[ k+i \ln\frac{t_1+\delta+e^{ik}t_2}{t_2+e^{ik}(t_1-\delta)} \right]\bigg|_{k=-\pi}^{k=\pi},
\end{eqnarray}
here, we need to ensure the winding can be well defined in Eq. (\ref{Windingsk}) which requires the denominator can not be zero along with the change of the momentum.
\begin{figure*}
\centering
\includegraphics[width=0.8\textwidth]{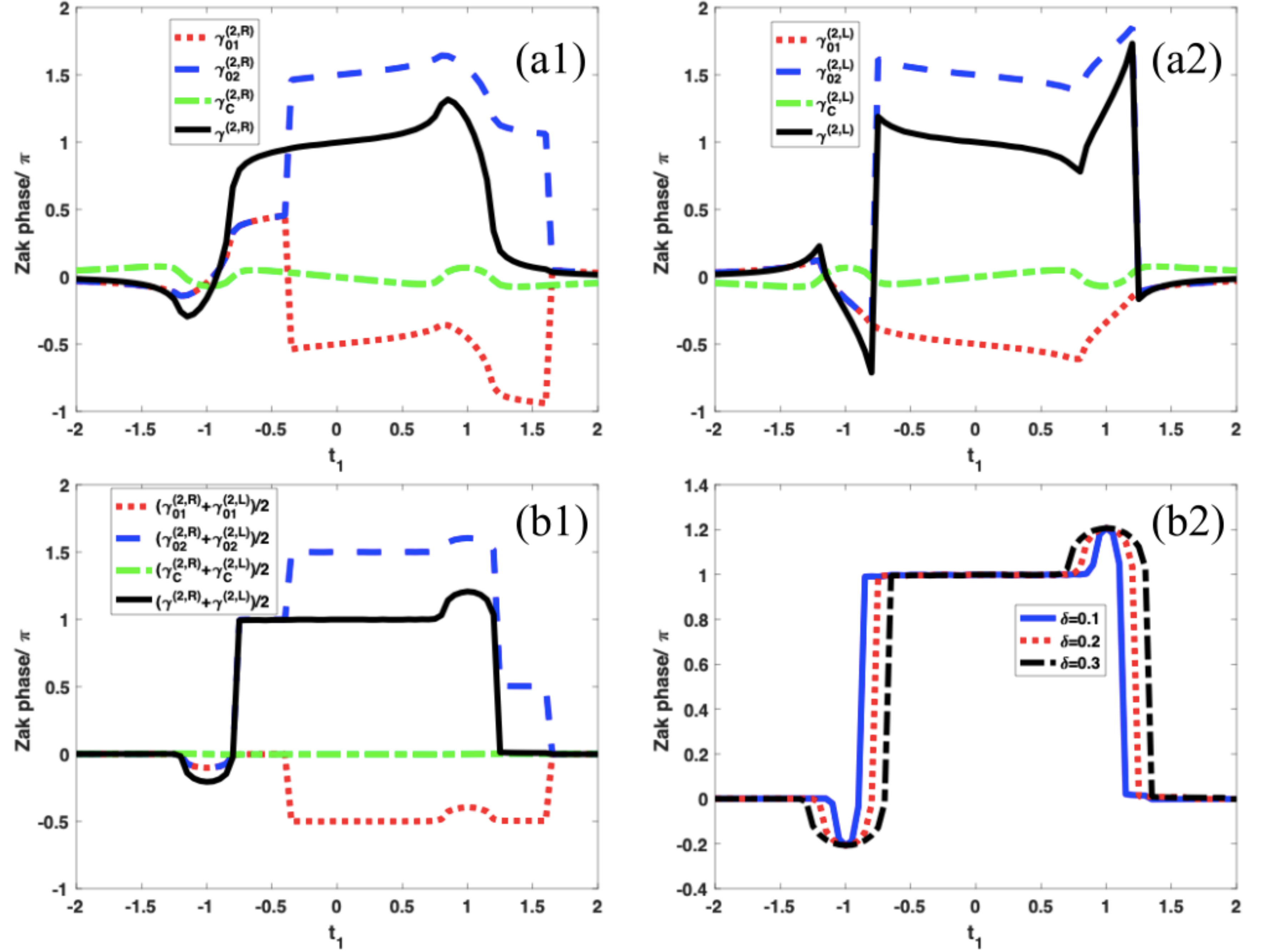}\\
\caption{The geometrical phases of the right eigenstate (a1) and the joint left eigenstate (a2) of $\psi_2$ with skin effect along with the change of $t_1$ under the Majorana's stellar representation. The mean values of each phase of the right eigenstates and the joint left eigenstates and the total phases of $\psi_2$ are shown in (b1) and (b2). We take $\delta$=0.2 in (a1), (a2), and (b1), and $t_2$=1 in all plots. }\label{skphase}
\end{figure*}

As is shown in Fig. \ref{skMSR}, the MSs of $\psi_1$ in Eq. (\ref{psisk}) are two fixed points $[1,0,0]$ and $[-1,0,0]$, however, stars of the MSR of $\psi_2$ and $\psi_3$ can move on the Bloch sphere along with the change of the momentum. We will take $\psi_2$ for example to study the geometrical phases of the right and the joint left eigenstates and how to define the winding number on the Bloch sphere. When $\left|t_1\right|>\left|t_2-\delta\right|$, the system is in the trivial phase and the winding number is zero calculated by Eq. (\ref{wingdingDF}). The same result can be found in Figs. \ref{skMSR} (a1) and (a2) which shows the MSR stars move around the singularity points and form a closed loop, but the singularity points are not included in the loop. When $\left|t_1\right|=0.9 t_2$, the MSR stars of the right eigenstates and the joint left eigenstates can't form a closed loop independently, however, they can form a loop that includes the singularity point by composition as is shown in Figs. \ref{skMSR} (b1) and (b2). But the system is still in the trivial phase  in this condition, because the normalized coefficient is a purely real number.  When $\left|t_1\right|$ gets smaller ($\left|t_1\right|<t_2-\delta$), each state can form the closed loop including the singularity points. At the same time, their combination loops will go around the $z$ axis  \cite{teo2020topological,jiang2018} as is illustrated in Figs. \ref{skMSR} (c1) and (c2).  

The MSR can give an intuitive way to observe the topological transition on the Bloch sphere, however, we still need to calculate the geometrical phases of the Hamiltonian with skin effect to investigate the correspondence between the MSR and the normal representation. In Fig. \ref{skphase}. The two independent phases and the correlation phases of $\psi_{2,R}$ and $\psi_{2,L}$ are shown in Figs. \ref{skphase}(a1)-(a2). All phases change continuously along with $t_1$, however, if we calculate the mean values of the phases of $\psi_{2,R}$ and $\psi_{2,L}$ on Bloch sphere, the topological transition can be recognized immediately as is vividly shown in Fig. (b1). When $\left|t_1\right|<t_2-\delta$ the total geometrical phase is exactly $\pi$ calculated by Eq. (\ref{HMphase}). The mean value of the correlation phase is always zero because of the normalized coefficient is a purely real number.

\section{The skin effect and the non-Bloch factor }\label{appendix2}
In this appendix, we will consider the real space Hamiltonian of Eq (\ref{skModel}). If the $\delta$ is zero, the Hamiltonian reduce to the conventional spin-1 Hermitian SSH model. If $\delta$ is beyond zero, we need to use the generalized Bloch theory to replace the Bloch factor $e^{ik}$ with $\beta=re^{ik}$ \cite{song2019}. The real space eigen-equation leads to
\begin{eqnarray}
E a_n&=&(t_1+\delta) b_n+t_2 b_{n+1}, \label{sk1}\\
E b_n&=&(t_1-\delta) a_n +t_2 a_{n-1}+(t_1-\delta) c_n+t_2 c_{n-1}, \\
E c_n&=&(t_1+\delta)\kappa_3+t_2 b_{n+1}, \label{sk3}
\end{eqnarray}
here, we can assume the shortcut solution with $(\phi_{n,A},\phi_{n,B},\phi_{n,C} )$=$\beta^n (\phi_{A},\phi_{B},\phi_{C} )$. If we substitute the assuming solution to Eqs. (\ref{sk1})- (\ref{sk3}), we can get $a_n$=$c_n$=$(t_1+\delta+t_2 \beta)b_n/E$ and the relation between $\beta$ and $E$ is $\beta=\frac{\pm\sqrt{\left(E^2-2 \left(-\delta ^2+t_1^2+t_2^2\right)\right)^2-16 t_2^2 (t_1-\delta ) (\delta +t_1)}+E^2-2 \left(-\delta ^2+t_1^2+t_2^2\right)}{4t_2(t_1-\delta )}$. Here, $\beta$ has two solution and in the $E\rightarrow0$ limit, we can find
\begin{equation}
\beta_{1,2}^{E\rightarrow0}=-\frac{t_2}{t_1-\delta}, -\frac{t_1+\delta}{t_2}.\label{betaEq}
\end{equation}
Comparing the solution between Eq. (\ref{betaEq}) and the Ref. \cite{song2019}, we can get the similar result of generalized Bloch factor
\begin{equation}
\beta=r=\sqrt{\left|\frac{t_1+\delta}{t_1-\delta}\right|}.
\end{equation}
It is obvious that the intracell hopping energy $\delta$ makes the Hamiltonian (\ref{skModel}) become non-Hermitian and breakdown the standard Bloch theory.
\begin{figure}
% \centering
\includegraphics[width=0.5\textwidth]{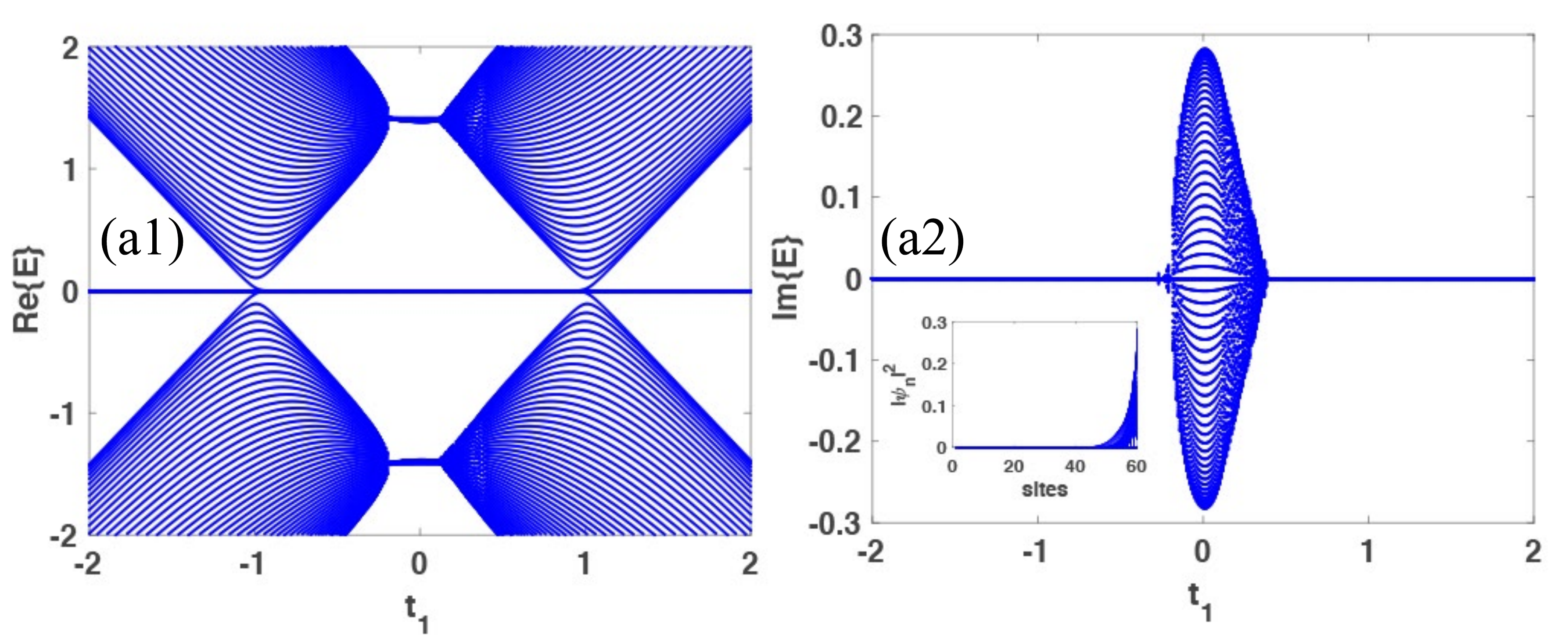}\\
\caption{The real parts (a1) and the imaginary parts (a2) of the eigen equations (\ref{sk1})-(\ref{sk3}) and the inset figure is the density distribution with $t_1$=0.8. We take $t_2$=1 and $\delta$=0.2. }\label{skband}
\end{figure}

The energy bands of Eqs. (\ref{sk1})-(\ref{sk3}) with the open boundary condition are illustrated in Fig. \ref{skband}. All the eigenstates are localized at the right side of the chain [see the inset picture in Fig. \ref{skband}(a2) ].

\end{document}